\documentclass[iop,apjl]{emulateapj}
\usepackage{apjfonts}
\usepackage{graphicx,natbib,hyperref}
\let\clearpage\relax

\newcommand\CaIIH{\mbox{Ca\,\textsc{ii}\,H}}
\newcommand\FeI{\mbox{Fe\,\textsc{i}}}

\newcommand\Ha{\mbox{H$\alpha$}}
\newcommand\CaII{\mbox{Ca\,\textsc{ii}}}

\begin{document}

\title{Probable identification of the on-disk counterpart of spicules in Hinode \CaIIH\ observations}
\shorttitle{Type-II spicules on disk}
\author{A.G.~de~Wijn}
\email{dwijn@ucar.edu}
\affil{High Altitude Observatory, National Center for Atmospheric Research\altaffilmark{1}, P.O. Box 3000, Boulder, CO 80307, USA}
\shortauthors{De Wijn}
\altaffiltext{1}{The National Center for Atmospheric Research is sponsored by the National Science Foundation.}

\begin{abstract}
I present a study of high-resolution time series of \CaIIH\ images and \FeI\ $630.15~\mathrm{nm}$ spectra taken with the Solar Optical Telescope on the Hinode spacecraft.
There is excellent correspondence between the \CaIIH\ and \FeI\ line core intensity, except tenuous emission around the network field concentrations in the former that is absent in the latter.
Analysis of on-disk observations and a comparison with limb observations suggests that this ``network haze'' corresponds to spicules, and likely to type-II spicules in particular.
They are known to appear in emission in on-disk broadband \CaIIH\ diagnostics
and the network haze is strongest in those areas where features similar to type-II spicules are produced in simulations.
\end{abstract}


\maketitle

\section{Introduction}\label{sec:introduction}

Spicules, mottles, and fibrils have long been studied as the principal component of the magnetized solar chromosphere.
The review by
	\cite{1968SoPh....3..367B}
on spicules at the limb and more recent work on mottles and fibrils on the disk
	\cite[e.g.,][]{2006ApJ...647L..73H,2007ApJ...655..624D}
all emphasize the crucial role of these features in the mass and energy balance of the solar chromosphere.
The relationship between spicules and similar features on disk has not been clearly established, largely because of the confusion resulting from projection as well as their fine structure and fast dynamics.

	\cite{2007PASJ...59S.655D}
argue that at least two species of spicules exist.
Both are tied intrinsically to magnetic field, but are different in their dynamics.
The traditional ``type-I'' spicules exhibit slower evolution ($3$--$7~\mathrm{min}$) and up-and-down motions with velocities on the order of $20~\mathrm{km}\,\mathrm{s}^{-1}$.
The ``type-II'' spicules, first identified as ``straws'' in near-limb \CaIIH\ images from the Dutch Open Telescope
	\citep{2004A&A...413.1183R}
by
	\cite{2006ASPC..354..276R,2007ASPC..368...27R},
	show much shorter lifetimes ($10$--$60~\mathrm{s}$), faster velocities ($50$--$150~\mathrm{km}\,\mathrm{s}^{-1}$), and often disappear over their whole length within a few seconds.
Further studies of type-II spicules have revealed that they are associated with Alfv\'enic waves
	\citep{2007Sci...318.1574D,2011Natur.475..477M}.

While ``straws'' or type-II spicules are clearly evident at or near the limb, they are not so readily identified on the disk.
Their faint, tenuous appearance makes them hard to discern against the background.
A search for disk counterparts by
	\cite{2008ApJ...679L.167L}
and a more detailed study by
	\cite{2009ApJ...705..272R}
revealed the existence of ``rapid blueshift excursions'' (RBEs) in the \Ha\ and \CaII\ IR lines that exhibit similar dynamics to type-II spicules and fit well with the finding by
	\cite{2009ApJ...701L...1D}
and
	\cite{2009ApJ...706L..80M}
that type-II spicules are possibly connected to the transition region and the corona through the blue-wing emission excess found in EUV lines by
	\cite{2008ApJ...678L..67H}
	\citep[see also][]{2010ApJ...718.1070H,
	2010ApJ...722.1013D,
	2011Sci...331...55D,
	2011ApJ...732...84M,
	2011ApJ...727L..37T,
	2011ApJ...738...18T,
	2012ApJ...746..158J}.

The correspondence of RBEs and type-II spicules seems well-established
	\citep{2012ApJ...752..108S}.
However, type-II spicules have been exclusively identified from Hinode images, and RBEs have only been identified in ground-based data from narrowband imaging instruments, so a direct identification of the on-disk counterpart of type-II spicules has yet to be presented.
In this Letter, I identify a weak haze around the magnetic network using Hinode data that I identify as a candidate for the on-disk manifestation of type-II spicules.

\section{Observations}\label{sec:observations}

\begin{figure*}[tbp]%
\begin{center}%
\includegraphics[width=180mm]{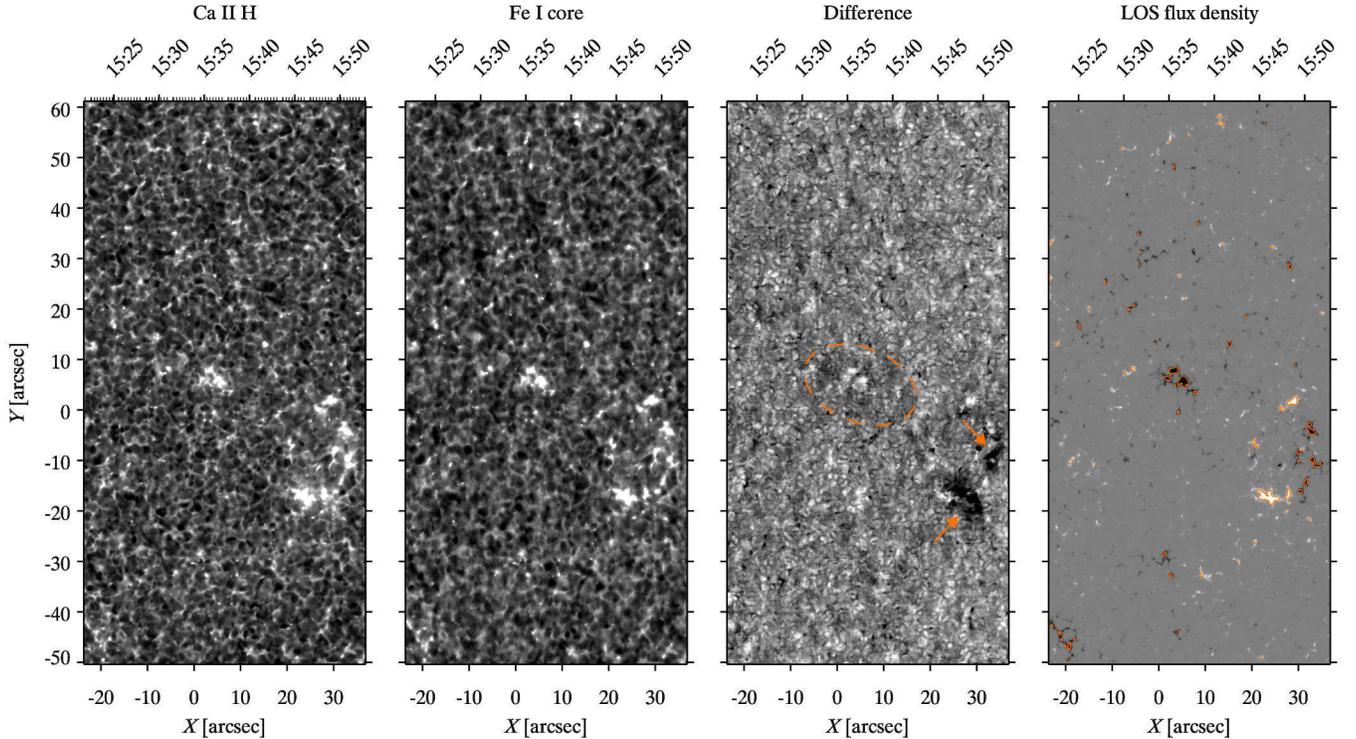}%
\caption{\CaIIH, \FeI\ $630.15~\mathrm{nm}$ core, \FeI\ $-$ \CaIIH\ difference image, and magnetic flux density images of the 19~April 2007 data set.
The SOT/SP observations were resampled in $(x,y)$ to match the SOT/BFI observations.
Time varies with $x$ to account for the scanning of the spectrograph.
The \CaIIH\ image sequence was resampled in time to match.
The observation time of individual \CaIIH\ images that make up the composite image is indicated by short tickmarks.
The difference image was made by linearly scaling the \FeI\ intensity to match the \CaIIH\ intensity in pixels that are between $10\arcsec$ and $14\arcsec$ distant from vertical flux greater than $500~\mathrm{Mx}\,\mathrm{cm}^{-2}$ (indicated by orange contours in the flux density image) and subtracting the \CaIIH\ intensity.
Dark and bright areas indicate an excess of \CaIIH\ and \FeI\ core intensity, respectively.
Two areas with strong \CaIIH\ brightness excess in the region of enhanced activity around ($28\arcsec$,$-18\arcsec$) are indicated with orange arrows, and a third more diffuse area is enclosed by a dashed line.
Flux concentrations themselves appear bright.
A faint signature of granulation is also visible, indicating that the reversed granulation in the \CaIIH\ filtergram has slightly higher contrast than in the \FeI\ core composite image.
The flux density image was scaled between $\pm500~\mathrm{Mx}\,\mathrm{cm}^{-2}$.}%
\label{fig:data_20070419}%
\end{center}%
\end{figure*}%

I employ observations from the Solar Optical Telescope
	\citep[SOT, ][]{2008SoPh..249..167T,2008SoPh..249..197S,2008SoPh..249..233I,2008SoPh..249..221S}
on board the \emph{Hinode} spacecraft
	\citep{2007SoPh..243....3K}.
The first sequence was taken on 19~April 2007 between 15:00 and 17:50~UT.
It contains G-band and \CaIIH\ image sequences and a map by the \emph{Hinode} SpectroPolarimeter (SP), both at the highest spatial resolution the instrument is capable of.
It covers an area of mostly quiet sun at disk center.
A small ephemeral region is present within the field of view during the sequence.
It was captured in the SP scan around 15:45~UT.
The second sequence was taken on 19~February 2007 between 00:15 and 6:01~UT.
This observation consists of a \CaIIH\ image sequence with $2\times2$ binning and an SP map at full resolution at the east limb.

I processed the observations with the standard SolarSoft \texttt{fg\_prep} and \texttt{sp\_prep} pipelines.
In addition, the SP observations were also processed with the Milne-Eddington inversion code MERLIN
	\citep{MERLIN}
to derive physical quantities such as field strength, filling factor, field azimuth, etc.
	\cite{2007ApJ...662L..31O}
found that such Milne-Eddington inversions are generally accurate in the case of SP observations for field strengths greater than $100~\mathrm{G}$
	\citep[cf.][]{2006A&A...456.1159M}.
Next, the SP observations were warped according to the model of
	\cite{2009ASPC..415..323C}
to align them with the BFI observations.
While the resulting alignment is already very good, I improved it for the 19~April 2007 data set by adjusting the parameters of the model in order to maximize the correlation between the SP \FeI\ continuum intensity and a G-band image interpolated in time at each spatial location to match the scanning of SP (henceforth referred to as a ``composite image'') using a cubic convolution method
	\citep{1983CGIP...23..258P}.
Then, I computed the plate scale difference and offset between the G-band and \CaIIH\ images by optimizing the correlation between a composite \CaIIH\ image and the \FeI\ $630.15~\mathrm{nm}$ line core intensity.
The Pearson correlation coefficient is $r=0.92$ for the image pair (shown in Fig.~\ref{fig:data_20070419}).
The G-band and SP observations were lastly resampled to match \CaIIH.

Visual inspection shows that the alignment between the data sets is very good.
To derive a quantitative measure of the quality of the alignment, I randomly chose 100 $64\times64~\mathrm{pixel}$ sub-fields and calculated the rigid offset between the \CaIIH\ and \FeI\ line core intensities.
I find an offset of $\Delta x=0.11\pm0.50~\mathrm{pixel}=(6\pm30)\times10^{-3}\arcsec$ and $\Delta y=-0.08\pm0.42~\mathrm{pixel}=(-4\pm23)\times10^{-3}\arcsec$.

The 19~February 2007 data set is more difficult to align as it does not include G-band filtergrams.
The \CaIIH\ and \FeI\ line core diagnostics exhibit different center-to-limb variation that makes them difficult to compare directly at the limb.
The parameters of the fit to the 19~April 2007 data set were used, but a new rigid alignment was calculated based on a subfield in the top-right corner of the field of view.
The correlation coefficient on the subfield is $r=0.74$, much lower than for the 19~April 2007 data.
However, visual inspection, e.g., of \CaIIH\ intensity and field concentrations from the \FeI\ inversions, shows that the co-alignment is excellent.

\section{Analysis}\label{sec:analysis}

\begin{figure}[tbp]%
\begin{center}%
\includegraphics[width=88mm]{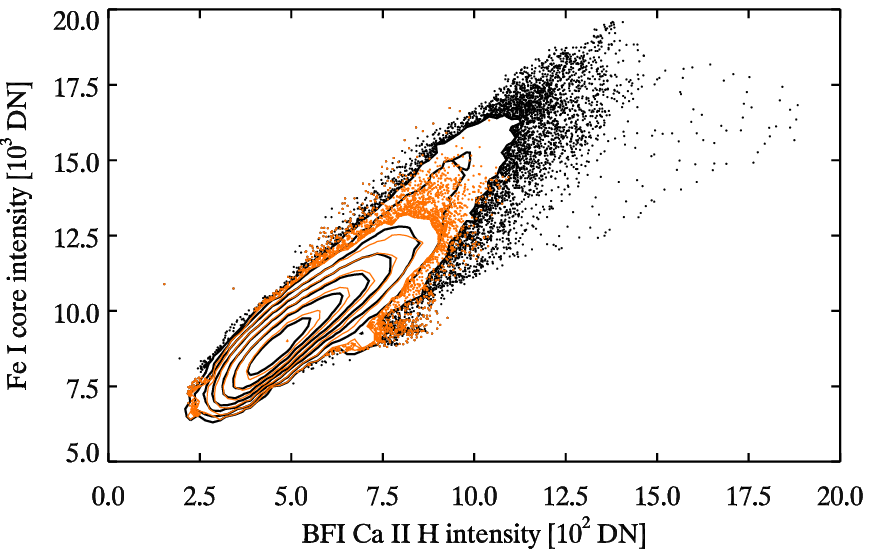}%
\caption{\CaIIH\ versus \FeI\ $630.15~\mathrm{nm}$ core intensities for all pixels (black) and for pixels between $1\arcsec$ and $5\arcsec$ from vertical magnetic flux $>500~\mathrm{Mx}\,\mathrm{cm}^{-2}$ (orange).}%
\label{fig:scat}%
\end{center}%
\end{figure}

Figure~\ref{fig:data_20070419} shows the \CaIIH\ composite image from BFI and \FeI\ $630.15~\mathrm{nm}$ line core from SP side-by-side.
The \FeI\ core image shows the same scene as the \CaIIH.
There is excellent correspondence between the \CaIIH\ and the \FeI\ core images ($r=0.92$, and see Fig.~\ref{fig:scat}).
However, the tenuous emission that surrounds the network in the \CaIIH\ image is absent in the \FeI\ image.
The region of emerging flux around ($20\arcsec$,$-15\arcsec$) shows this network haze most clearly, but it is also faintly visible around network vertices, e.g., at ($5\arcsec$,$5\arcsec$).
The third panel shows a ``difference image'', in which the \CaIIH\ intensity is subtracted from the \FeI\ core intensity.
The \FeI\ core intensity was scaled linearly in order to match the \CaIIH\ intensity on pixels that are between $10\arcsec$ and $14\arcsec$ removed from areas of vertical flux greater than $500~\mathrm{Mx}\,\mathrm{cm}^{-2}$ (indicated by orange contours in the rightmost panel of Fig.~\ref{fig:data_20070419}).
The difference image shows a clear excess of \CaIIH\ emission around the emerging flux region (black areas indicated by arrows around ($30\arcsec$,$-20\arcsec$)).
A more diffuse area of \CaIIH\ brightness excess surrounds the network vertex at ($5\arcsec$,$5\arcsec$).
Flux concentrations themselves, on the other hand, are brighter in the \FeI\ core.
A faint granular pattern also emerges, indicating that the reversed granulation in the \CaIIH\ filtergram has slightly higher contrast than in the \FeI\ core composite image.

\begin{figure}[tbp]%
\begin{center}%
\includegraphics[width=88mm]{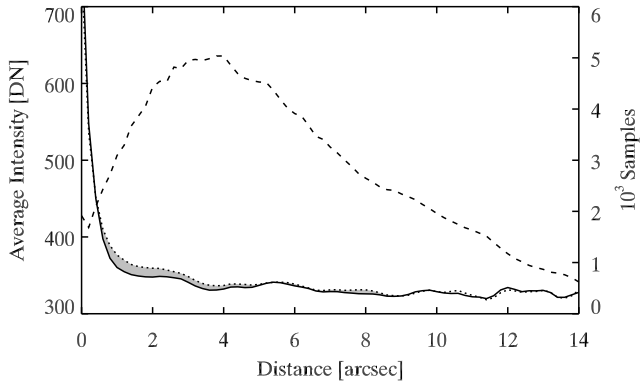}%
\caption{\FeI\ $630.15~\mathrm{nm}$ core (solid line) and \CaIIH\ (dotted line) average intensity and pixel counts (dashed line) in $0.2\arcsec$ bins as a function of distance to vertical magnetic flux $>500~\mathrm{Mx}\,\mathrm{cm}^{-2}$ (orange contours in the rightmost panel of Fig.~\ref{fig:data_20070419}).
All pixels in the field of view are considered.
The \FeI\ core intensity was scaled linearly to match the \CaIIH\ intensity over the range $10$--$14\arcsec$.
The ``network haze'' is most evident between $1\arcsec$ and $5\arcsec$ as a brightness excess in the \CaIIH\ intensity (shaded area).}%
\label{fig:haze}%
\end{center}%
\end{figure}%

\begin{figure}[tbp]%
\begin{center}%
\includegraphics[width=88mm]{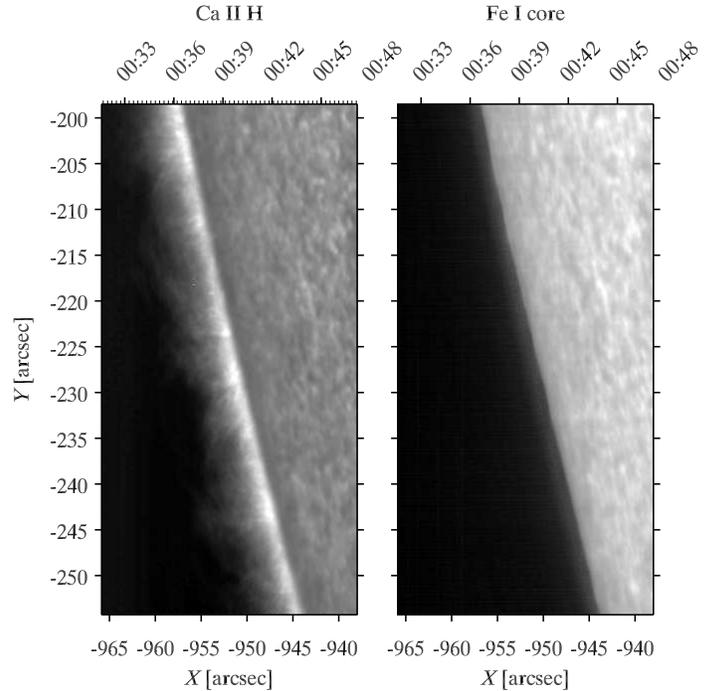}%
\caption{Limb cutout of the \CaIIH\ and \FeI\ $630.15~\mathrm{nm}$ core images of the 19~February 2007 data set, in the same format as Fig.~\ref{fig:data_20070419}.
A radial filter was applied to enhance structures on the limb in both images.}%
\label{fig:data_20070219}%
\end{center}%
\end{figure}%

The \CaIIH\ brightness excess is concentrated around locations of strong vertical magnetic flux.
In Fig.~\ref{fig:haze} I plot the intensity as a function of distance from regions with vertical flux greater than $500~\mathrm{Mx}\,\mathrm{cm}^{-2}$.
The \FeI\ core was scaled in the same way as in the difference image in Fig.~\ref{fig:data_20070419}.
The network haze is evident as a brightness excess in the \CaIIH\ intensity of several percent over the intensity in the \FeI\ core at a distance between $1\arcsec$ and $5\arcsec$ from areas of strong vertical flux ($37\%$ of all pixels).
The photometric noise on the images is small and can be estimated with the SolarSoft \texttt{fg\_noise} function.
I find about $3\times10^{-4}$ for both the \CaIIH\ images as well as the \FeI\ core composite image.
For the latter, this number is consistent with the typical estimate of $10^3$ polarimetric signal-to-noise for an SP ``normal map''.
In addition, the curves in Fig.~\ref{fig:haze} average over several thousand samples.

Figure~\ref{fig:scat} shows a scatterplot of \CaIIH\ versus \FeI\ core intensities for both all pixels and those pixels between $0\arcsec$ and $5\arcsec$ removed from strong vertical magnetic flux.
The two clouds are homomorphic, indicating that the pixels close to strong vertical magnetic flux do not constitute a distinct population.
However, the difference image shows that most of the excess is concentrated around the emerging flux region, and closer inspection shows that the bulge in the scatterplot around $(7,9)$ is associated with these pixels.
This shows that network haze is exhibited by pixels with average intensity in the SOT \CaIIH\ passband but slightly below average intensity in the \FeI\ core.

Figure~\ref{fig:data_20070219} shows the comparison between \CaIIH\ and \FeI\ core intensity at the limb using the data set from 19~February 2007.
In this case, SOT only observed \CaIIH\ with the BFI and \FeI\ with the SP.
A radial filter was applied to both the \CaIIH\ and \FeI\ data to scale the brightness of the off-limb structures by a factor of $5$.
Both type-I and type-II spicules are visible in \CaIIH\ as a thick forest above the limb that reaches to $5\arcsec$ height.
The \FeI\ core intensity image shows very faint emission up to about $1\arcsec$ above the limb that was studied in detail by
	\cite{2010ApJ...713..450L}.

\section{Discussion}\label{sec:discussion}

There are three contributions to \CaIIH\ network brightness in observations with relatively broad filter passbands such as these that cause an excess over the internetwork brightness.
Because the filter is broad compared to the core of the \CaIIH\ line the majority of the intensity in these observations is photospheric sampled by the line wings.
The classic Wilson depression allows emission to escape from deeper and hotter layers and hence the brightness of the wings is increased
	\citep[e.g.,][]{2005A&A...437.1069S}.
Calculations of the response function by
	\cite{2007PASJ...59S.663C}
show that the passband also samples higher layers.
Heating in the magnetic network in the high photosphere and low chromosphere thus adds to the network brightness excess
	\citep[e.g.,][]{2009A&A...503..577C}.
Finally, there is a contribution from chromospheric structures in the core of the \CaIIH\ line
	\citep[][]{2009A&A...500.1239R}.

The core of the \FeI\ line samples the same atmospheric layers as the Hinode \CaIIH\ passband as is evidenced by the exceedingly high correlation coefficient between the two diagnostics.
As a result, the first two sources of brightness excess in the \CaIIH\ images will also contribute in the same way to the excess of brightness in the network as seen in the core of the \FeI\ line.
The third source, however, is mostly absent in the \FeI\ observations because of the lack of optical thickness of spicules in that line.
A weak emission signal extending less than $1\arcsec$ above the limb was recently identified by
	\cite{2010ApJ...713..450L}.
They analyze the emission and conclude that the polarization signature is most likely created by scattering and that it is depolarized through the Hanle effect by a magnetic field of several Gauss.

Limb observations show that the distinct \CaIIH\ spicule forest is not present in the \FeI\ core intensity.
This, and the correspondence of the length of the spicules in \CaIIH\ to the extent of the on-disk network haze (see Fig.~\ref{fig:haze}), suggests that the haze is caused by spicules.

The persistence of some arch-like structures on the limb is noteworthy.
The chromospheric magnetic field, stable on timescales longer than a few minutes, is apparently traced out by spicules preferentially along some field lines.

Type-I spicules have been extensively studied in \CaIIH\
	\citep{1968SoPh....3..367B}.
Mottles and fibrils, on the other hand, have not been identified in \CaIIH\ diagnostics.
They have been studied in chromospheric observations in lines such as \Ha\ and the IR \CaII\ triplet.
It stands to reason that they must have some signature also in the \CaIIH\ line, but they probably appear as absorption features that would be difficult to detect against the background of reversed granulation.

It is more likely that the network haze is the on-disk counterpart of type-II spicules.
Type-II spicules have been detected on disk as emission ``straws'' by
	\cite{2006ASPC..354..276R,2007ASPC..368...27R}.
They cannot be individually identified in these \CaIIH\ images (neither in the composite images shown in Figs.~\ref{fig:data_20070419} and~\ref{fig:data_20070219} nor in the individual images of which they are made up).
This can be attributed to several factors.
First, the fast dynamics of the very slender type-II spicules are not captured in these \CaIIH\ images because they are resampled in time to match the slow scanning of the SP slit.
Furthermore, the wide SOT \CaIIH\ bandpass of $0.3~\mathrm{nm}$ samples too much light in the wings of the line for the type-II spicules to show up with high contrast in disk observations.
It is possible to observe type-II spicules directly on the limb with SOT because there is no bright background
	\citep[e.g.,][]{2007PASJ...59S.655D},
and even near the limb
	\citep{2010ApJ...714L...1S},
though in both cases there is confusion from integration along the line of sight.
In SOT \CaIIH\ filtergrams, type-II spicules would be very faint on disk, yet must have some emission signature.
RBEs, on the other hand, appear in absorption against the photosphere in \Ha\ because that line has no opacity in the region of the atmosphere where reversed granulation is formed
	\citep{2006A&A...449.1209L}.

The \CaIIH\ intensity excess is strongest in the region of emerging flux round ($20\arcsec$,$-15\arcsec$).
It is however also present and faintly visible around other areas of concentrated flux, e.g., the network vertex at ($0\arcsec$,$5\arcsec$).
The average intensity as a function of distance to strong flux concentrations still shows excess \CaIIH\ emission between $1\arcsec$ and $5\arcsec$ even if the emerging flux region is excluded from the analysis.
This is in agreement with the results of
	\cite{2012ApJ...752..108S}
who studied RBEs and found they are ubiquitous but concentrated around regions of enhanced magnetic field.
From simple tests it appears that the results presented here are robust with respect to the choice of threshold value for the selection of concentrations of strong vertical flux.
If the region of emerging flux is excluded, the \FeI\ intensity shows an excess over the \CaIIH\ intensity for distances less than $1\arcsec$ from strong vertical flux, consistent with the observation in Fig.~\ref{fig:data_20070419} that the flux concentrations themselves are bright in the difference image.

The simulations by
	\cite{2011ApJ...736....9M}
also provide evidence to support the identification of the haze with type-II spicules.
They find structures in their realistic 3D MHD simulations that resemble type-II spicules in predominantly unipolar regions of emerging flux.
The difference image in Fig.~\ref{fig:data_20070419} shows that the network haze is strongest in the two patches of predominantly unipolar flux of opposite polarity in the region of enhanced activity around ($28\arcsec$,$-18\arcsec$) and ($33\arcsec$,$-10\arcsec$), in agreement with the simulations.
The objects in the simulation are produced as a result of deflection of plasma that is pushed against a ``wall'' of low-$\beta$ plasma by a strong Lorentz force.
The regions of emerging flux exhibit the strong gradients in the magnetic field in both their simulation and these observations, which is required for the mechanism to work effectively.
Ideally, for comparison with these simulations one would not only determine the locations of network haze but also the history of flows in these areas through feature tracking on a sequence of magnetograms.
The observations used here unfortunately did not include such a sequence.

Because of the relatively broad SOT \CaIIH\ filter, the wings of the line dominate the contribution to the reversed granulation intensity.
If a diagnostic can be found that samples similar structures as the \CaIIH\ wing but that lacks opacity in chromospheric structures, it may be possible to increase the contrast of spicules on disk through subtraction.
Such a procedure has previously been used for the identification of magnetic bright points
	\citep{1998ApJ...509..435V}.
The \FeI\ line core would appear to be a suitable candidate.
However, the filter bandwidth must be narrow.
The SP samples the \FeI\ line very cleanly with a spectral resolution of $3.0~\mathrm{pm}$ sampled with $2.15~\mathrm{pm}$ pixels.
The difference image in Fig.~\ref{fig:data_20070219} suffers from the spatial resolution of the SP data as well as the resampling of the \CaIIH\ data in time to match slit scanning.

Summarizing, there is evidence to suggest that the network haze in the \CaIIH\ images is the on-disk manifestation of some class spicules.
Type-II spicules are the likely candidate, since they are known to appear in emission in on-disk \CaIIH\ images from other other telescopes
	\citep{2006ASPC..354..276R,2007ASPC..368...27R},
and the region around which is appears most clearly is where type-II spicules are expected to be produced based on simulations
	\citep{2011ApJ...736....9M}.

\acknowledgments{I am indebted to R.~J.~Rutten for discussions and comments.
\emph{Hinode} is a Japanese mission developed and launched by ISAS/JAXA, with NAOJ as domestic partner and NASA and STFC (UK) as international partners. It is operated by these agencies in co-operation with ESA and NSC (Norway).
Hinode SOT/SP Inversions were conducted at NCAR under the framework of the Community Spectro-polarimtetric Analysis Center (CSAC; \url{http://www.csac.hao.ucar.edu/}).
This research made use of NASA's Astrophysics Data System.}

\end{document}